\documentclass{aastex}
\usepackage{spr-astr-addons}
\usepackage{url}

\RequirePackage{color}

\begin{document}

\title{Modified Chaplygin Gas and Solvable F-essence Cosmologies}

\shorttitle{Solvable F-essence Cosmologies} \shortauthors{Myrzakulov
et al.}

\author{Mubasher Jamil\altaffilmark{1}}
\and
\author{Yerlan Myrzakulov\altaffilmark{2}}
\and
\author{Olga Razina\altaffilmark{2}}
\and
\author{Ratbay Myrzakulov\altaffilmark{2,}\altaffilmark{3}}

\altaffiltext{1}{Center for Advanced Mathematics and Physics,
National University of Sciences and Technology, Islamabad, Pakistan.
Email: mjamil@camp.nust.edu.pk}

\altaffiltext{2}{Eurasian International Center for Theoretical
Physics, Eurasian National University, Astana 010008, Kazakhstan}

\altaffiltext{3}{Department of Physics, California State University,
Fresno, CA 93740 USA, Email: rmyrzakulov@csufresno.edu ,
rmyrzakulov@gmail.com}

\begin{abstract}

The Modified Chaplygin Gas (MCG) model belongs to the class of a
unified models of dark energy and dark matter. In this paper, we
have modeled MCG in the framework of f-essence cosmology. By
constructing an equation connecting the MCG and the f-essence, we
solve it to obtain explicitly the pressure and energy density of
MCG. As special cases, we obtain both positive and negative pressure
solutions for suitable choices of free parameters. We also calculate
the state parameter which describes the phantom crossing.

\end{abstract}

\sloppy

\section{Introduction}

Several complementary cosmological observations guide us that our
Universe is experiencing an accelerated expansion in the current era
\citep{perl,ries}. From the observations of Wilkinson Microwave
Anisotropy Probe (WMAP) satellite which gathered data about the
cosmic microwave background radiation, such a cosmic acceleration is
produced by a so-called dark energy (DE) \citep{2}. Such a new
element of the Universe, capable of accelerating, must be, in
accordance with the Friedmann equation, have a pressure less than
minus one third of the energy density. The notion of a cosmic medium
possessing negative pressure is not new in cosmology. The very first
proposal for dark energy is the cosmological constant denoted by
$\Lambda$. Originally proposed by Einstein to construct a
theoretical model of the universe such that the spherical
configuration of matter in the universe is balanced with the
negative pressure of cosmological constant. Thereby creating a
static universe, in the conformity of the notions held by many
scientists of that time. The origin of cosmic acceleration in recent
time needs $\Lambda$ but it faces serious problems of fine tuning
(the value of $\Lambda$ is several orders of magnitude larger then
estimated from the empirical results) and cosmic coincidence problem
(the energy density of matter and dark energy component are
approximately same in our presence). Then theorists looked for other
candidates of dark energy. There has been a wide variety of
theoretical models of dark energy constructed in the literature
including quintessence, phantom energy, Chaplygin gas, tachyon and
dilaton dark energy etc, see \citep{rev} for further details. The
quintessence and phantom energy models are based on spatially
homogeneous and time dependent scalar fields, in conformity of the
cosmological principle. The Lagrangians for the quintessence
(phantom energy) has positive (negative) kinetic energy. Where
quintessence faces still fine tuning problem of the parameters of
its potential, while the phantom energy gives very esoteric
possibilities of Big Rip and black hole evaporations \citep{qadir}.

Quintessence as a model of dark energy relies on the suitable choice
of the potential function or the potential energy of scalar fields.
It is also possible that the cosmic acceleration could appear due to
modification of the kinetic energy of the scalar fields. Such
modifications are termed non-canonical. The kinetically driven
cosmic acceleration was originally proposed as a model for
inflation, namely k-inflation \citep{Mukhanov1}, and then as a model
for dark energy, namely k-essence
\citep{Mukhanov2,Mukhanov3,Chiba,Chiba1,Scherrer, yang,Linder,capo,karami,khodam,adabi,umar}.
This model is free from fine-tuning and anthropic arguments.
K-essence has been proposed as a possible means of explaining the
coincidence problem of the Universe beginning to accelerate only at
the present epoch \citep{cope}. Instead, k-essence is based on the
idea of a dynamical attractor solution which causes it to act as a
cosmological constant only at the onset of matter-domination.
Consequently, k-essence overtakes the matter density and induces
cosmic acceleration at about the present epoch. In some models of
k-essence, the cosmic acceleration continues forever while in
others, it continues for a finite duration \citep{Mukhanov3}.

 In the last years, the k-essence model has received much attention.
It is still worth investigating in a systematic way the possible
cosmological behavior of the k-essence.  Quite recently, a model
named g-essence is proposed \citep{MR} which is a more generalized
version of k-essence. In fact, the g-essence contains, as particular
cases, two important models: k-essence and f-essence. Note that
f-essence is the fermionic counterpart of k-essence.

To our knowledge, in the literature there are relatively few works
on dark energy models with fermionic fields. However, in the recent
years several approaches were made to explain the accelerated
expansion by taking fermionic fields as the gravitational sources of
energy (see e.g. refs.
\citep{Ribas1,Kremer01,Kremer02,MR0,Tsyba,MR3,Kremer03,Cai,Wang,
Ribas2,Rakhi1,Rakhi2,Chimento,Anischenko,Saha1,Saha2,Saha3,
Saha4,Saha5,Vakili1,Wei,Dereli,Balantekin1,Armendariz-Picon}). In
particular, it was shown that the fermionic field plays very
important role in: (i) isotropization of initially anisotropic
spacetime; (ii) formation of singularity free cosmological
solutions; (iii) explaining late-time acceleration.

A very appealing proposal to describe the dark sector are the
so-called unified models. The prototype of such model is the
Chaplygin gas. In the unified models, dark energy and dark matter
are described by a single fluid, which behaves as ordinary matter in
the past, and as a cosmological constant term in the future. In this
sense, it interpolates the different periods of evolution of the
Universe, including the present state of accelerated expansion. The
Chaplygin gas model leads to very good results when confronted with
the observational data of supernova type Ia. Concerning the matter
power spectrum data, the statistic analysis leads to results
competitive with the $\Lambda$CDM model, but the unified (called
quartessence) scenario must be imposed from the beginning. It means
that the only pressureless component is the usual baryonic one,
otherwise there is a conflict between the constraints obtained from
the matter power spectrum and the supernova tests. Note that many
variations of the Chaplygin gas model have been proposed in the
literature. One of them is the so-called Modified Chaplygin Gas
model.  It is important that the MCG model belongs to the class of a
unified models of dark energy and dark matter. In this context, it
is important to study the relation between MCG and the other unified
models of dark energy and dark matter. For example, in
\citep{MRKessence} relationship between MCG and k-essence was
established. In this paper, we have modeled MCG in the framework of
f-essence cosmology. By constructing an equation connecting the MCG
and the f-essence, we solve it to obtain explicitly the pressure and
energy density of MCG. As special cases, we obtain both positive and
negative pressure solutions for suitable choices of free parameters.

This paper is organized as follows. In section II, we introduce the
F-essence formalism. In section III, we briefly discuss Modified
Chaplygin gas and its connection with the f-essence. In section IV,
we construct a governing differential equation of our model and
solve it for several special cases and a general case. Conclusion is
presented in the last section. In the Appendix we provide
 the derivation of the equations of motion of g-essence, k-essence and f-essence.

\section{F-essence}

Let us briefly present some basics of f-essence. Its action has the
form \citep{MR0,Tsyba}
\begin{equation}
S=\int d^{4}x\sqrt{-g}[R+2K(Y, \psi, \bar{\psi})],
\end{equation}
where $\psi$ and $\bar{\psi}=\psi^{\dagger}\gamma^0$ denote the
fermion field and its adjoint, respectively, the dagger represents
complex conjugation and $R$ is the Ricci scalar. The fermionic
fields are treated here as classically commuting fields (see e.g.
refs. \citep{Ribas1, Cai,
Wang,Rakhi2,Chimento,Saha1,Vakili1,Wei,Dereli,Balantekin1,Armendariz-Picon}).
What we mean by a classical fermionic field is a set of four
complex-valued space-time functions that transform according to the
spinor representation of the Lorenz group. The existence of such
fields is crucial in our work since despite the fact that fermions
are described by quantized fermionic fields which do not have a
classical limit, we assume such classical fields exist and use them
as matter fields coupled to gravity. A possible justification for
the existence of classical fermionic fields is given in the appendix
of reference \citep{Armendariz-Picon}. So for a more extensive and
physically detailed discussion of the properties of such classical
fermionic fields we refer to this fundamental work
\citep{Armendariz-Picon}. Furthermore, $K$ is the Lagrangian density
of the fermionic field, the canonical kinetic term has the form
\begin{equation}
Y=\frac{1}{2}i[\bar{\psi}\Gamma^{\mu}D_{\mu}\psi-(D_{\mu}\bar{\psi})\Gamma^{\mu}\psi].
\end{equation}
Moreover, $\Gamma^{\mu}=e^{\mu}_a\gamma^a$ are the generalized Dirac-Pauli
matrices satisfying the Clifford algebra
\begin{equation}
\{\gamma^\mu, \gamma^\nu\}=2g^{\mu\nu},
\end{equation}
where the braces denote the anti-commutation relation. $e^{\mu}_a$ denotes
the tetrad or "vierbein"  while the covariant derivatives are given by
\begin{equation}
D_\mu\psi=\partial_\mu\psi-\Omega_\mu\psi,\quad D_\mu\bar{\psi}=\partial_\mu\bar{\psi}+\bar{\psi}\Omega_\mu.
\end{equation}
Above, the fermionic connection $\Omega_\mu$ is defined by
\begin{equation}
\Omega_\mu=-\frac{1}{4}g_{\rho\sigma}[\Gamma^{\rho}_{\mu\delta}-e^{\rho}_{b}\partial_\mu e^{b}_{\delta}]\Gamma^{\sigma}\Gamma^{\delta},
\end{equation}
with $\Gamma^{\rho}_{\mu\delta}$ denoting the Christoffel symbols.

We work with the Friedmann-Robertson-Walker (FRW) spacetime given by
\begin{equation}
ds^2=-dt^2+a^2(dx^2+dy^2+dz^2),
\end{equation}
For this metric, the vierbein is chosen to be
\begin{equation}
(e_a^\mu)=diag(1,1/a,1/a,1/a),\quad (e^a_\mu)=diag(1,a,a,a).
\end{equation}
The Dirac matrices of curved spacetime $\Gamma^\mu$ are
\begin{equation}
\Gamma^0=\gamma^0,   \Gamma^j=a^{-1}\gamma^j,   \Gamma^5=-i\sqrt{-g}\Gamma^0\Gamma^1\Gamma^2\Gamma^3=\gamma^5,
\end{equation}
\begin{equation}
\Gamma_0=\gamma^0,  \quad \Gamma_j=a\gamma^j, \quad (j=1,2,3).
\end{equation}
Hence we get
\begin{equation}
\Omega_0=0, \quad  \Omega_j=0.5\dot{a}\gamma^j\gamma^0
\end{equation}
and $$
Y=0.5i(\bar{\psi}\gamma^0\dot{\psi}-\dot{\bar{\psi}}\gamma^0\psi).
$$
 Finally, we note that the gamma matrices we write in the Dirac basis that is as
 \begin{equation}
\gamma^0 = \begin{pmatrix} I & 0 \\ 0 & -I \end{pmatrix}, \quad\gamma^k = \begin{pmatrix} 0 & \sigma^k \\ -\sigma^k & 0 \end{pmatrix},\quad \gamma^5 = \begin{pmatrix} 0 & I \\ I & 0 \end{pmatrix},
\end{equation}
where $I=diag (1,1)$ and the $\sigma^k$ are Pauli matrices having the following form
 \begin{equation}
\sigma^1 = \begin{pmatrix} 0 & 1 \\ 1 & 0 \end{pmatrix},\quad \sigma^2 = \begin{pmatrix} 0 & -i \\ i & 0 \end{pmatrix},\quad \sigma^3 = \begin{pmatrix} 1 & 0 \\ 0 & -1 \end{pmatrix}.
\end{equation}
 We are ready to write  the equations of f-essence (In detail, the derivation of these equation  we will  present in the Appendix). Here we write the final form of these equations. We have
\begin{eqnarray}
3H^2-\rho&=&0,\\
2\dot{H}+3H^2+p&=&0,\\
K_{Y}\dot{\psi}+0.5(3HK_{Y}+\dot{K}_{Y})\psi-i\gamma^0K_{\bar{\psi}}&=&0,\\
K_{Y}\dot{\bar{\psi}}+0.5(3HK_{Y}+\dot{K}_{Y})\bar{\psi}+iK_{\psi}\gamma^{0}&=&0,\\
\dot{\rho}+3H(\rho+p)&=&0,
\end{eqnarray}
where
\begin{equation}
\rho=YK_{Y}-K,\quad p=K,
\end{equation}
are the energy density and pressure of the fermionic field.  If
$K=Y-V,$ then from the system (13)-(18) we get the corresponding
equations of  the Einstein-Dirac model. At last, we note that f-essence is the particular
 case of g-essence for  which the   energy density and pressure
are given by
\begin{equation}
\rho=2XK_X+YK_{Y}-K,\quad p=K,
\end{equation}
where $X=0.5\dot{\phi}^2$ is the canonical kinetic term for the
scalar field $\phi$ (see the Appendix).

\section{Modified Chaplygin gas and f-essence}

In the cosmological context, the Chaplygin gas was first suggested
as an alternative to quintessence and was demonstrated to have an
increasing $\Lambda$ (cosmological constant) behavior for the
evolution of the universe \citep{Kamenshchik,bento}. The EoS of the
MCG dark energy model was proposed by Benaoum \citep{Benaoum} as an
exotic fluid which could explain the cosmic accelerated expansion.
Later on, it was shown that the EoS of MCG is valid from the
radiation era to $\Lambda$CDM model \citep{ujjal}. The MCG
parameters $\alpha$ and $B$ have been constrained by the cosmic
microwave background CMB data \citep{li}. The stable scaling
solutions (attractor) of the Freidmann equation have been obtained
in \citep{stable,stable1}. The MCG is given by \citep{Benaoum}
\begin{equation}
p=A\rho-\frac{B}{\rho^\alpha},
\end{equation}
where $A$ and $B$ are positive constants and $0\leq\alpha\leq 1.$ In
the light of $3-$year WMAP and the SDSS data, the MCG best fits the
data by choosing $A=-0.085$ and $\alpha=1.724$ \citep{lu}. The
dynamical attractor for the MCG exists at $\omega=-1$ (where
$p=\omega\rho$), hence MCG can do the phantom crossing from
$\omega>-1$ to $\omega<-1$, independent to the choice of initial
conditions \citep{jing}. A generalization of MCG was proposed in
\citep{debnath} by taking $B\equiv B(a)=B_0 a^{n}$, where $n$ and
$B_0$ are constants. The MCG is the generalization of generalized
Chaplygin gas $p=-B/\rho^\alpha$ \citep{sen,carturan} with the
addition of a barotropic term. Using Eqs. (17) and (20), we can show that the
MCG energy density and pressure are given by
\citep{Benaoum}
\begin{eqnarray}
\rho&=&\Big[B(1+A)^{-1}+Ca^{-3(1+\alpha)(1+A)}\Big]^{\frac{1}{1+\alpha}},\\
p&=&[AB(1+A)^{-1}+ACa^{-3(1+\alpha)(1+A)}]-\nonumber\\&&B\Big[B(1+A)^{-1}
+Ca^{-3(1+\alpha)(1+A)}\Big]^{-\frac{\alpha}{1+\alpha}},
\end{eqnarray}
where $C$ is a constant of integration. This constant of integration $C$
can be found from the condition that the fluid has vanishing pressure $p=0$ when $a=a_0$:
$$
C=B[A(1+A)]^{-1}a_0^{3(1+\alpha)(1+A)}.
$$
For the modified Chaplygin gas  case,  the EoS parameter is
\begin{equation}
\omega=\frac{ACa^{-3(1+\alpha)(1+A)}-B(1+A)^{-1}}{B(1+A)^{-1}+Ca^{-3(1+\alpha)(1+B)}}.
\end{equation}
From (18) we get
\begin{eqnarray}
\rho&=&\frac{dK}{d\ln Y}-K,\\
p&=&K.
\end{eqnarray}
Hence follows that
$$
\rho+p=\frac{dp}{d\ln Y}=\frac{dp}{da}\frac{da}{d\ln Y}
$$
so that  we have
\begin{equation}
\frac{d\ln Y}{da}=\frac{p_a}{p+\rho},
\end{equation}
where $p_a=\frac{dp}{da}$. The last equation has the following solution
\begin{equation}
 Y=Y_0e^{\int\frac{p_a}{p+\rho}da},
\end{equation}
where $Y_0$ is an integration constant.  We can rewrite this expression as
    \begin{equation}
 Y=Y_0e^{\int\frac{p_\zeta}{p+\rho}d\zeta},
\end{equation}
where $\zeta=Ca^{-3(1+\alpha)(1+A)}.$ Then the  corresponding expressions for the energy density and pressure take the form
\begin{eqnarray}
\rho&=&\Big[D+\zeta\Big]^{\frac{1}{1+\alpha}},\\
p&=&[A(D+\zeta)-B]\Big[D+\zeta\Big]^{-\frac{\alpha}{1+\alpha}}
\end{eqnarray}
with $D=B(1+A)^{-1}$. Then we have
\begin{eqnarray}
p+\rho&=&(1+A)\zeta(D+\zeta)^{-\frac{\alpha}{1+\alpha}},\\
p_\zeta&=&A(D+\zeta)^{-\frac{\alpha}{1+\alpha}}\nonumber\\&&-\frac{\alpha}{1+\alpha}
(A\zeta-D)(D+\zeta)^{-\frac{\alpha}{1+\alpha}-1}.
\end{eqnarray}
Hence we get
\begin{equation}
\int\frac{p_\zeta}{p+\rho}d\zeta=\ln\Big[\zeta^{\frac{(1+\alpha)A+\alpha}
{(1+\alpha)(1+A)}}(D+\zeta)^{-\frac{\alpha}{1+\alpha}}\Big].
\end{equation}
Finally we have
\begin{equation}
Y=Y_0\zeta^{\frac{(1+\alpha)A+\alpha}{(1+\alpha)(1+A)}}(D+\zeta)
^{-\frac{\alpha}{1+\alpha}}.
\end{equation}

\section{Solvable f-essence cosmologies }
In this section, we consider some solvable f-essence models related
 with the MCG given by  (20).
\subsection{The case: $B=0$}In this case the EoS takes the form
 \begin{equation}
p=A\rho,
\end{equation}
where $\rho$ evolves like
 \begin{equation}
\rho=\rho_0a^{-3(1+A)},\quad \rho_0=const
\end{equation}
so that
 \begin{equation}
p=A\rho_0a^{-3(1+A)}.
\end{equation}
\begin{figure}
\centering
\includegraphics[scale=.9]{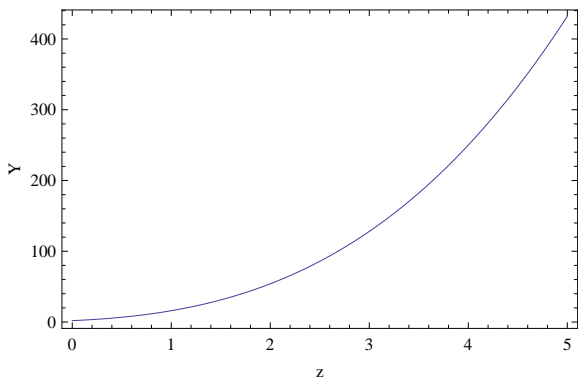}
\caption{ Evolution of kinetic energy $Y$ as a function of redshift
$z$ from Eq. (40). Other parameters are fixed at $Y_0=2$ and
$A=-1$.}
\end{figure}
On the other hand, the equations (24)-(25) for  (35) give
 \begin{equation}
\rho=FY^{\frac{1+A}{A}},\quad F=const.
\end{equation}
In this case, the pressure is
 \begin{equation}
p\equiv K=AFY^{\frac{1+A}{A}}.
\end{equation}
Comparison of (36) and (38) yields
\begin{equation}
Y=Y_0a^{-3A},\quad Y_0=\rho_0^{\frac{A}{1+A}}F^{-\frac{A}{1+A}}.
\end{equation}
We get the scale factor as
\begin{equation}
a(t)=\Big(\frac{Y}{Y_0}\Big)^{\frac{-1}{3A}}.
\end{equation}
The behavior of Eq. (40) against redshift $a^{-1}-1=z$ is plotted in
figure-1 where we see that this evolution is of power-law form.
\begin{figure}
\centering
 \includegraphics[scale=.9]{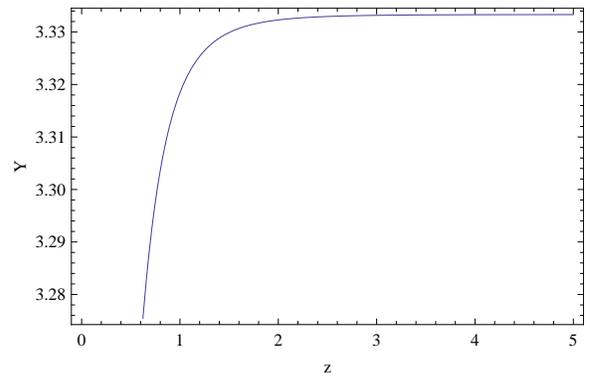}
  \caption{Evolution of $Y$ against $z$ from Eq. (47). Other parameters are
  fixed at $C=0.5$, $B=0.4$, $W=0.3$ and $\alpha=1.2$ }
\end{figure}
\subsection{The case: $A=0$}
Ignoring the barotropic term in MCG, we have
 \begin{equation}
p=-\frac{B}{\rho^\alpha}.
\end{equation}
It is called the generalized Chaplygin gas (GCG)
\citep{Kamenshchik}. Recently it is proposed using perturbative
analysis and power spectrum observational data that the MCG model is
not a successful candidate for the cosmic medium unless $A=0$, i.e.
the usual GCG model is favored \citep{f}. As well-known, the
corresponding energy density and pressure are given by
\begin{eqnarray}
\rho&=&\Big[B+Ca^{-3(1+\alpha)}\Big]^{\frac{1}{1+\alpha}},\\
p&=&-B\Big[B+Ca^{-3(1+\alpha)}\Big]^{-\frac{\alpha}{1+\alpha}},
\end{eqnarray}
where $C$ is a constant of integration. From (34), (43) and (44) we
get the expressions for the energy density and pressure:
\begin{equation}
\rho=\Big[\frac{B}{1-(WY)^{\frac{1+\alpha}{\alpha}}}\Big]^{\frac{1}{1+\alpha}},
\end{equation}
\begin{equation}
p=-B^{\frac{1}{1+\alpha}}\Big[1-(WY)^{\frac{1+\alpha}
{\alpha}}\Big]^{\frac{\alpha}{1+\alpha}},
\end{equation}
where $W=const$.  The solution for $Y$ is determined from (43) and
(45):
\begin{equation}
Y=W^{-1}C^{\frac{\alpha}{1+\alpha}}\Big[C+Ba^{3(1+\alpha)}
\Big]^{-\frac{\alpha}{1+\alpha}}.
\end{equation}

The behavior of Eq. (47) is shown in figure-2, where we see that the
kinetic energy of the f-essence increases and then stays constant at
higher redshifts. Note that this conclusion depends crucially on the
chosen values of free parameters.

\subsection{The general case}

In this section, we consider the general case when $A\neq 0,\quad
B\neq 0$. So in this case we must solve the following system
\begin{eqnarray}
\rho&=&YK_{Y}-K,\\
K&=&A\rho-\frac{B}{\rho^\alpha}
\end{eqnarray}
or
\begin{eqnarray}
\rho&=&Yp_{Y}-p,\\
p&=&A\rho-\frac{B}{\rho^\alpha}.
\end{eqnarray}
Solving equation (48) for $K$, we arrive at
\begin{equation}
p\equiv K=EY+Y\int\frac{\rho}{Y^2}dY,
\end{equation}
where $E$ is an integration  constant. Note that if
$\rho=V(\bar{\psi}, \psi)$ then from (52), it follows that
$K=EY-V(\bar{\psi}, \psi)$ i.e. the purely Dirac case. Eqs. (51) and
(52) give
\begin{equation}
EY+Y\int\frac{\rho}{Y^2}dY=A\rho-\frac{B}{\rho^\alpha},
\end{equation}
which has the following solution:
\begin{equation}
(1+A)\rho^{1+\alpha}-(WY)^{\frac{n(1+\alpha)}{\alpha}}\rho^{n(1+\alpha)}-B=0,
\end{equation}
where $W$ is a constant  and
 \begin{equation}
n=\frac{\alpha(1+A)}{A+\alpha(1+A)}.
\end{equation}
From (55) it follows that $A$ and $\alpha$ are related by
 \begin{equation}
A=-\frac{(n-1)\alpha}{(n-1)\alpha+n} \quad or \quad \alpha=-\frac{nA}{(n-1)(1+A)}.
\end{equation}
The search of the analytical solutions of Eq. (54) is a tough job.
So let us find some particular solutions for some values of $n$.

\subsubsection{Example 1: $n=0$}

It follows from (55) that this case ($n=0$)  realized as $\alpha=0$ or
$A=-1$.
\begin{figure}
\centering
\includegraphics[scale=.9]{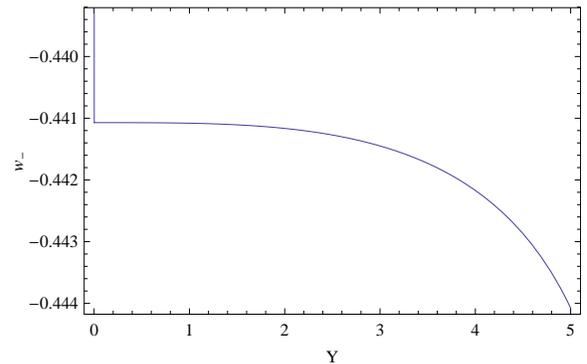}
\caption{For $n=2$, the EoS parameter $w_-$ is plotted against the
kinetic energy $Y$, while other parameters are fixed at $B=0.1$,
$W=0.2$, $\alpha=1.5$ }
\end{figure}
1) Let us first consider the case $\alpha=0$. Then $n=0$ and the
equations (48)-(49)  take the form
\begin{eqnarray}
    \rho&=&YK_{Y}-K,\\
        K&=&A\rho-B
    \end{eqnarray}
    and the equation (54) becomes\begin{equation}
(1+A)\rho^{1+\alpha}-(WY)^{\frac{1+A}{A}}-B=0.
\end{equation}
Hence we write
\begin{equation}
\rho=(1+A)^{-1}[B+(WY)^{\frac{1+A}{A}}].
\end{equation}
and for the pressure
\begin{equation}
p=(1+A)^{-1}[-B+A(WY)^{\frac{1+A}{A}}].
\end{equation}
The corresponding equation of state  (EoS) parameter is given by
\begin{equation}
\omega=A-\frac{B(1+A)}{B+(WY)^{\frac{1+A}{A}}}.
\end{equation}

2) Now we consider the case when  $A=-1.$  Then $n=0$ and equations
(48)-(49)  take the form
\begin{eqnarray}
\rho&=&YK_{Y}-K,\\
K&=&-\rho-B\rho^{-\alpha}.
\end{eqnarray}
Hence we get
\begin{equation}
\alpha B\ln\rho-(1+\alpha)^{-1}\rho^{1+\alpha}=\ln (C_1Y)^{-B}.
\end{equation}
Consider some particular solutions  of this equation. If $\alpha=0$, then we have
\begin{equation}
\rho=\ln (C_1Y)^{B},
\end{equation}
and for the pressure
\begin{equation}
p=\ln (C_1Y)^{-B}-B.
\end{equation}
The corresponding EoS parameter is given by
\begin{equation}
\omega=-1-[\ln (C_1Y)]^{-1}.
\end{equation}
Second example is $\alpha=-1$. Then for the energy density and pressure we obtain
\begin{equation}
\rho=C_2Y^{\frac{B}{1+B}}\quad (C_2=const),
\end{equation}
and\begin{figure}
\centering
\includegraphics[scale=.9]{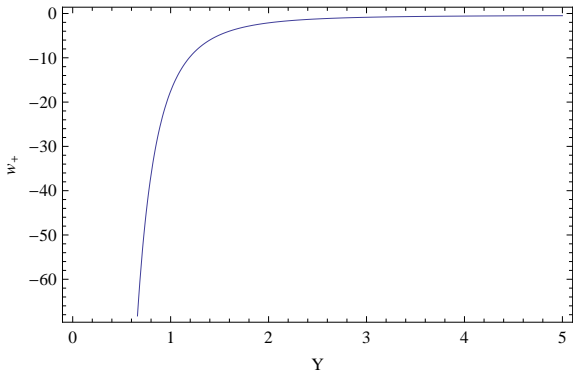}
\caption{ For $n=2$, the EoS parameter $w_+$ is plotted against the
kinetic energy $Y$, while other parameters are fixed at $B=0.1$,
$W=0.2$, $\alpha=1.5$}
\end{figure}
\begin{equation}
p=-(1+B)C_2Y^{\frac{B}{1+B}}.
\end{equation}
The corresponding EoS parameter is given by
\begin{equation}
\omega=-1-B.
\end{equation}

\subsubsection{Example 2: $n=1$}

If $n=1$ then from (56), it follows that $A=0$. This case was
considered in subsection 4.2,  so we omit it here.

\subsubsection{Example 3: $n=2$}

\begin{figure}
\centering
\includegraphics[scale=.9]{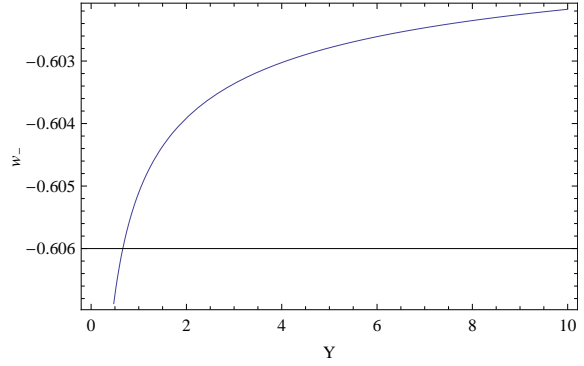}
\caption{For $n=0.5$, the EoS parameter $w_-$ is plotted against the
kinetic energy $Y$, while other parameters are fixed at $B=-0.1$,
$W=0.2$, $\alpha=-1.5$ }
\end{figure}
Now we consider the case when $n=2$. In this case $A$ and $\alpha$
related by
\begin{equation}
A=-\frac{\alpha}{\alpha+2} \quad or\quad
\alpha=-\frac{2A}{1+A}.
\end{equation}
The equation for $\rho$ (54) takes the form
\begin{equation}
(1+A)\rho^{1+\alpha}-(WY)^{\frac{2(1+\alpha)}{\alpha}}\rho^{2(1+\alpha)}-B=0.
\end{equation}
It has the solution
\begin{eqnarray}
\rho&=&(WY)^{-\frac{2}{\alpha}}\Big\{\frac{1}{1+\alpha}
\nonumber\\&&\times\Big[1\mp\sqrt{1-B(1+\alpha)^2(WY)^{\frac{2(1+\alpha)}
{\alpha}}}\Big]\Big\}^{\frac{1}{1+\alpha}}.
\end{eqnarray}
The pressure is given by
\begin{eqnarray}
p&=&-\frac{\alpha(WY)^{-\frac{2}{\alpha}}}{\alpha+2}
\Big\{\frac{1}{1+\alpha}\nonumber\\&&\times\Big[1\mp\sqrt{1-B(1+\alpha)
^2(WY)^{\frac{2(1+\alpha)}{\alpha}}}\Big]\Big\}^{\frac{1}{1+\alpha}}\nonumber\\&&-
B(WY)^{2}\Big\{\frac{1}{1+\alpha}\nonumber\\&&\times\Big[1\mp\sqrt{1-
B(1+\alpha)^2(WY)^{\frac{2(1+\alpha)}{\alpha}}}\Big]\Big\}
^{-\frac{\alpha}{1+\alpha}}.
\end{eqnarray}\begin{figure}
\centering
\includegraphics[scale=.9]{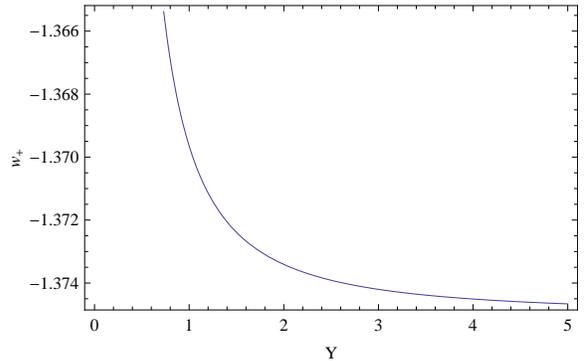}
\caption{For $n=-1$, the EoS parameter $w_+$ is plotted against the
kinetic energy $Y$, while other parameters are fixed at $B=-1$,
$W=2$, $\alpha=1.5$ }
\end{figure}
The corresponding EoS parameter is
\begin{eqnarray}
\omega_\mp&=&-\frac{\alpha}{\alpha+2}-B(WY)^{-\frac{2(1+\alpha)}{\alpha}}\Big\{
\frac{1}{1+\alpha}\nonumber\\&&\times\Big[1\mp\sqrt{1-
B(1+\alpha)^2(WY)^{\frac{2(1+\alpha)}{\alpha}}}\Big]\Big\}.
\end{eqnarray}
In figures (3) and (4), we have plotted the EoS parameter and it is
shown that subnegative values of $\omega$ are permissible in our
model. This corresponds to f-essence MCG behaving as phantom energy
which causes super-accelerated expansion \citep{ujjal}.

\subsubsection{Example 4: $n=0.5$}

If $n=0.5$ then $A$ and $\alpha$ satisfy the relation
\begin{equation}
A=-\frac{\alpha}{\alpha-1} \quad or\quad
\alpha=\frac{A}{1+A}.
\end{equation}
Eq. (54) becomes
\begin{equation}
(1+A)\rho^{1+\alpha}-(WY)^{\frac{(1+\alpha)}{2\alpha}}\rho^{0.5(1+\alpha)}-B=0.
\end{equation}
This equation  has the following solution
\begin{equation}
\rho=(WY)^{\frac{1}{\alpha}}\Big\{\frac{1-\alpha}{2}
\Big[1\pm\sqrt{1+\frac{4B}{1-\alpha}(WY)^{-\frac{1+\alpha}
{\alpha}}}\Big]\Big\}^{\frac{2}{1+\alpha}}.
\end{equation}
The pressure is given by
\begin{eqnarray}
p&=&-\frac{\alpha(WY)^{\frac{1}{\alpha}}}{\alpha-1}\Big\{\frac{1-\alpha}{2}
\nonumber\\&&\times
\Big[1\pm\sqrt{1+\frac{4B}{1-\alpha}(WY)^{-\frac{1+\alpha}
{\alpha}}}\Big]\Big\}^{\frac{2}{1+\alpha}}\nonumber\\&&-
B(WY)^{-1}\Big\{\frac{1-\alpha}{2}\nonumber\\&&\times
\Big[1\pm\sqrt{1+\frac{4B}{1-\alpha}(WY)^{-\frac{1+\alpha}
{\alpha}}}\Big]\Big\}^{-\frac{2\alpha}{1+\alpha}}.
\end{eqnarray}\begin{figure}
\centering
\includegraphics[scale=.9]{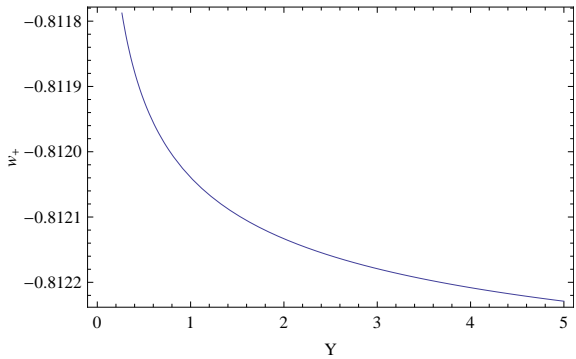}
\caption{ For $n=-1$, the EoS parameter $w_+$ is plotted against the
kinetic energy $Y$, while other parameters are fixed at $B=-3$,
$W=0.8$, $\alpha=-1.5$ }
\end{figure}
The corresponding EoS parameter reads
\begin{eqnarray}
\omega_\mp&=&-\frac{\alpha}{\alpha-1}-
\frac{(WY)^{\frac{1+\alpha}{\alpha}}}{1-\alpha}\nonumber\\&&\times
\Big[1\mp\sqrt{1+\frac{4B}{1-\alpha}(WY)^{-\frac{1+\alpha}
{\alpha}}}\Big]^{2}.
\end{eqnarray}
In figure-5, we have plotted the EoS parameter against the kinetic
term. Here we chose $\alpha<0$ which corresponds to the polytropic
term added in the barotropic equation of state. Such a f-essence
polytropic EoS can also cause the super-acceleration.

\subsubsection{Example 5: $n=-1$}

Our next example is  $n=-1$. Then $A$ and $\alpha$ satisfy the relation
\begin{equation}
A=-\frac{2\alpha}{2\alpha-1} \quad or\quad
\alpha=\frac{A}{2(1+A)}.
\end{equation}
Eq. (54) becomes
\begin{equation}
(1+A)\rho^{1+\alpha}-(WY)^{-\frac{1+\alpha}{\alpha}}\rho^{-(1+\alpha)}-B=0.
\end{equation}
It has the solution
\begin{equation}
\rho=\Big[\frac{B\pm\sqrt{B^2+4(1+A)(WY)^{-\frac{1+\alpha}
{\alpha}}}}{2(1+A)}\Big]^{\frac{1}{1+\alpha}}.
\end{equation}
The pressure is given by
\begin{eqnarray}
p&=&A\Big[\frac{B\pm\sqrt{B^2+4(1+A)(WY)^{-\frac{1+\alpha}{\alpha}}}}
{2(1+A)}\Big]^{\frac{1}{1+\alpha}}\nonumber\\&&-B\Big[\frac{B\pm\sqrt{B^2+4(1+A)(WY)
^{-\frac{1+\alpha}{\alpha}}}}
{2(1+A)}\Big]^{-\frac{\alpha}{1+\alpha}}.
\end{eqnarray}
The corresponding EoS parameter is given by
\begin{eqnarray}
\omega_\mp&=&-\frac{2\alpha}{2\alpha-1}+\frac{B}{2}(WY)^{\frac{1+\alpha}
{\alpha}}\nonumber\\&&\times\Big[B\mp\sqrt{B^2-\frac{1}{2\alpha-1}(WY)
^{-\frac{1+\alpha}{\alpha}}}\Big].
\end{eqnarray}
In figures 6 and 7, we plotted the above state parameter against
kinetic energy. We choose positive and negative values of $\alpha$
for the sake of completeness. It is apparent that the state
parameter achieves subnegative values showing the viability of our
dark energy model.

\section{Conclusion}

In summary, we have modeled modified Chaplygin gas in f-essence
cosmology. The use of MCG is useful as a tool of explaining dark
energy and dark matter in a unified manner, while f-essence
cosmology essentially suitable to describe cosmic acceleration only
at present time. Thus the correspondence of MCG with f-essence is
useful in learning how these two models are connected to each other.
We studied this link by constructing a differential equation (54)
connecting the MCG and the f-essence. We solved it to obtain
explicitly the pressure and energy density of MCG. We observed that
f-essence MCG has one additional free parameter namely $W$ along
with $A$, $B$, $\alpha$. As special cases, we obtain both positive
(37), (80) and negative (46), (68), (72) pressure solutions for
suitable choices of free parameters. The negative pressure solution
is essentially useful for cosmic expansion with acceleration. Prior
to this, we studied the model with barotropic and generalized
Chaplygin gas equation of states.

The present work is concerned only with the correspondence of MCG
with the f-essence. However it would be more interesting to study
the dynamical features of this model. For instance, it is
interesting to check whether such a model is useful to model
inflation at earlier cosmic epoch and cosmic acceleration at the
present time. Next it will be important to compare this model with
the observational data and constrain its free parameters,
particularly $W$. Also observational constraints on the state
parameter $\omega$ will indicate if it corresponds to cosmological
constant, quintessence or phantom energy. We finish our work in
the hope that merits and demerits of f-essence will be clear in the future investigations.

We note that this paper is the logical continuation of
\citep{MRKessence}, where the relation between k-essence and MCG was
studied. The action of k-essence reads as
\begin{equation}
S=\int d^{4}x\sqrt{-g}[R+2K(X, \phi)],
\end{equation}
where the kinetic term $X$ for the scalar field $\phi$ (for the FRW metric) reads as
\begin{equation}\label{1y}
X=0.5\dot{\phi}^2.
\end{equation}
The corresponding equations for the FRW metric look like
\begin{eqnarray}\label{1t}
    3H^2-\rho &=&0,\\
        2\dot{H}+3H^2+p&=&0,\\
        K_{X}\ddot{\phi}+(\dot{K}_{X}+3HK_{X})\dot{\phi}-K_{\phi}&=&0,\label{1w}\\
            \dot{\rho}+3H(\rho+p)&=&0,\label{2w}
    \end{eqnarray} where the energy density and pressure of the scalar  field is given by
\begin{equation}
\rho=2XK_{X}-K,\quad p=K.
\end{equation}
In this case, the common system of equations for the k-essence and MCG has the form
\begin{eqnarray}\label{1001}
    \rho&=&2Xp_{X}-p,\\ \label{1002}
        p&=&A\rho-\frac{B}{\rho^\alpha}.
    \end{eqnarray}
The compatibility condition for the equations
\eqref{1001}-\eqref{1002} is given by \citep{MRKessence}
\begin{equation}\label{1b}
(1+A)\rho^{1+\alpha}-(WX)^{\frac{n(1+\alpha)}{2\alpha}}\rho^{n(1+\alpha)}-B=0,
\end{equation}
where $W=W(\phi)$  and
 \begin{equation}\label{1n}
n=\frac{\alpha(1+A)}{A+\alpha(1+A)}.
\end{equation}
In \citep{MRKessence}, different type  solvable k-essence
cosmologies compatible with the MCG model are found for the
 different values of $n$.

\section*{Acknowledgments}
We would like to thank the anonymous referees for providing us with constructive comments and suggestions to improve this work. One of the authors (R.M.) thanks   D. Singleton  and Department of Physics, California State University Fresno  for their hospitality during his one year visit (October, 2010 -- October, 2011).

 \section{APPENDIX. The derivation of the equations of motion of g-essence, k-essence and f-essence}
In this Appendix we would like  to present the derivation of the equations of motion for  the f-essence action (2.1) that is the system (13)-(17).  But, as f-essence is the exact particular case of g-essence, we  consider more general case and give the derivation of the equations of motion for g-essence. Let us consider the following action of g-essence
\begin {equation} \label{}
S=\int d^{4}x\sqrt{-g}[R+2K(X, Y,  \phi, \psi, \bar{\psi})],
\end{equation}
where $R$ is the scalar curvature,   $X$ is
the kinetic term for the scalar field $\phi$, $Y$ is the kinetic term for the fermionic field $\psi$ and  $K$ is some function (Lagrangian) of its arguments. In the case of  the FRW  metric
\begin{equation} \label{}
ds^2=-dt^2+a^2(dx^2+dy^2+dz^2),
\end{equation}
 $R$, $X$ and $Y$ have the form
\begin{equation} \label{}
 R=6\left(\frac{\ddot{a}}{a}+\frac{\dot{a}^2}{a^2}\right),
\end{equation}
\begin{equation} \label{}
X=0.5\dot{\phi}^2,
\end{equation}
\begin{equation} \label{}
Y=0.5i(\bar{\psi}\gamma^0\dot{\psi}-\dot{\bar{\psi}}\gamma^0\psi),
\end{equation}
respectively.
 Substituting (100)-(102)  into (98) and integrating over the spatial dimensions,
 we are led to an effective Lagrangian in the mini-superspace $\{a,\phi, \psi, \bar{\psi}\}$
 \begin{equation}
L=-2(3a\dot{a}^2-a^3K).\end{equation}
 Variation of Lagrangian (103) with respect to $a$ yields the equation of motion
 of the scale factor
 \begin{equation}
    2a\ddot{a}+\dot{a}^2+a^2K=0.
\end{equation}
 Now by varying the above Lagrangian (103) with respect to the scalar field $\phi$ we
 obtain its equation of motion as
 \begin{equation}
K_X\ddot{\phi}+\dot{K}_X\dot{\phi}+3\frac{\dot{a}}{a}K_X \dot{\phi}-K_{\phi}=0.
\end{equation}
At least, the variation of Lagrangian (103) with respect to $\bar{\psi}, \psi$ that is the
 corresponding Euler-Lagrangian equations for the fermionic fields give
\begin{equation}
K_Y\gamma^0\dot{\psi}+1.5\frac{\dot{a}}{a}K_Y\gamma^0\psi+0.5\dot{K}_Y\gamma^0\psi-iK_{\bar{\psi}}=0,
\end{equation}
\begin{equation}
K_Y\dot{\bar{\psi}}\gamma^0+1.5\frac{\dot{a}}{a}K_Y\bar{\psi}\gamma^0+0.5\dot{K}_Y\bar{\psi}\gamma^0+iK_{\psi}=0.
\end{equation}
Also, we have the "zero-energy" condition given by
\begin{equation}
L_{\dot{a}}\dot{a}+L_{\dot{\phi}}\dot{\phi}+L_{\dot{\psi}}\dot{\psi}+L_{\dot{\bar{\psi}}}\dot{\bar{\psi}}-L=0
\end{equation}
which yields the constraint equation
\begin{equation}
-3a^{-2}\dot{a}^2+2XK_X+YK_Y-K=0.
\end{equation}
Collecting all derived equations (104) - (107) and (109) and rewriting them using the Hubble parameter $H=(\ln a)_t,$ we come to the following closed system of equations of g-essence (for the FRW metric case):
\begin{eqnarray}
3H^2-\rho&=&0,\\
2\dot{H}+3H^2+p&=&0,\\
K_X\ddot{\phi}+(\dot{K}_X+3HK_X) \dot{\phi}-K_{\phi}&=&0,\\
K_{Y}\dot{\psi}+0.5(3HK_{Y}+\dot{K}_{Y})\psi-i\gamma^0K_{\bar{\psi}}&=&0,\\
K_{Y}\dot{\bar{\psi}}+0.5(3HK_{Y}+\dot{K}_{Y})\bar{\psi}+iK_{\psi}\gamma^{0}&=&0,\\
\dot{\rho}+3H(\rho+p)&=&0.
\end{eqnarray}
Here
\begin{equation}
\rho=2XK_X+YK_{Y}-K,\quad p=K
\end{equation}
are the energy density and pressure of g-essence. It is clear that these expressions for the energy density and pressure represent the components of the energy-momentum tensor of g-essence:
\begin{equation}
T_{00}=2XK_X+YK_{Y}-K,\quad  T_{11}=T_{22}=T_{33}=-K.
\end{equation}
Also we note that g-essence admits two particular cases (reductions): k-essence and f-essence. In fact, let $\psi=0$. Then $Y=0$ and the system (110)-(115) becomes
\begin{eqnarray}
3H^2-\rho&=&0,\\
2\dot{H}+3H^2+p&=&0,\\
K_X\ddot{\phi}+(\dot{K}_X+3HK_X) \dot{\phi}-K_{\phi}&=&0,\\
\dot{\rho}+3H(\rho+p)&=&0,
\end{eqnarray}
where
\begin{equation}
\rho=2XK_{X}-K,\quad p=K.
\end{equation}
It is the system of equations of k-essence. Now let us consider the case
 when $\phi=0$ that is the purely fermionic case. Then $X=0$, $K=K(Y,
  \bar{\psi}, \psi)$ and the system (110)-(115) takes the form of the
  equations of f-essence that is (13)-(17).

\end{document}